\documentclass[10pt,letterpaper]{article}
\usepackage{opex3}

\begin{document}

%%%%%%%%%%%%%%%%%% title page information %%%%%%%%%%%%%%%%%%
\title{Tight focusing of plane waves from micro-fabricated spherical mirrors}

\author{J. Goldwin and E. A. Hinds}

\address{Centre for Cold Matter,\\
Department of Physics, Division of Quantum Optics and Laser Science,\\
Imperial College, London SW7 2AZ, United Kingdom}

\email{j.goldwin@imperial.ac.uk} %% email address is required

\homepage{http://www3.imperial.ac.uk/ccm} %% author's URL, if desired


\begin{abstract}
We derive a formula for the light field of a monochromatic plane
wave that is truncated and reflected by a spherical mirror. Within
the scalar field approximation, our formula is valid even for deep
mirrors, where the aperture radius approaches the radius of
curvature. We apply this result to micro-fabricated mirrors whose
size scales are in the range of tens to hundreds of wavelengths,
and show that sub-wavelength focusing (full-width at half-maximum
intensity) can be achieved.  This opens up the possibility of
scalable arrays of tightly focused optical dipole traps without
the need for high-performance optical systems.
\end{abstract}


%%%%%%%%%%%%%%%%%%%%%%% References %%%%%%%%%%%%%%%%%%%%%%%%%

%%%%%%%%%%%%%%%%%%%%%%%%%%  body  %%%%%%%%%%%%%%%%%%%%%%%%%%
\section{Introduction}

Spherical mirrors are widely used, for example in the telescope,
to collect light from a distant object and focus it to a point. In
practice this produces not a point but a distribution of light
over a finite region in the vicinity of the focus. Analysis of
this behavior involves two classic topics in optics ---
diffraction from a circular aperture and spherical aberrations of
the mirror. Typically, the spot size is characterized by a figure
of merit such as transverse aberration or root-mean-squared blur
radius. However, in a variety of modern applications, such as data
storage, optical tweezers or atom trapping, it is only important
to have a narrow central spot even though this may be accompanied
by a broad distribution of low intensity in the wings. For such
applications, the analysis requires a new approach, which we
develop here.

Our own interest is in making an array of atom traps so small that
each trap can hold one atom (or none), but not two. If a second
atom is caught, the two interact through a light-induced,
inelastic collision and both are kicked out of the trap --- an
effect known as collisional blockade
\cite{Schlosser01,Schlosser02}. The ejection of atom pairs occurs
promptly provided the spot size of the trapping light is below
$\sim 1 \mu$m \cite{Sortais07}, which is roughly the size of the
optical wavelength $\lambda$. One important application for such
traps is the production of single photons on demand
\cite{Darquie05,Weber06,Tey08}, which can then be used as a
powerful resource for quantum cryptography and quantum information
processing \cite{Lounis05,Waks02,Knill01,Oxborrow05}.

There is a growing literature concerned with strong focusing of
light onto single atoms in free space, as a means of
achieving efficient atom-light coupling.  A scheme proposed in
Ref.~\cite{Sondermann07} aims to emulate the field radiated from a
single atom. This ideal can be approached using a parabolic
reflector to focus light with highly sculpted intensity and phase
profiles, but the scheme does not seem suitable for scaling to
large arrays of traps. In Ref.~\cite{vanEnk01}, van Enk and Kimble
considered the simpler case of focusing a uniformly polarized
Gaussian beam with a lens that imposes a Gaussian phase profile.
They were able to achieve a three-dimensional solution for the
light field in the region of the focal point. Tey \emph{et al.}
recently extended this analysis to the case of a spherically
converging Gaussian wave \cite{Tey08b}, which gave somewhat better
coupling. Both of these calculations neglect the aberrations
produced by a real imaging system.

In the work presented here, we consider focusing by spherical
mirrors and we treat the aberrations of the mirror exactly. No
high-performance optical system is required and the method offers
a simple way to achieve large arrays of traps that are tight
enough to produce a strong collisional blockade for alkali atoms.
The idea is to use arrays of hemispherical mirrors, typically with
radius of curvature $R\sim100\,\mu$m, that are lithographically
etched directly into the surface of a silicon wafer
\cite{Eriksson05,Trupke05} and covered with a reflective coating.
An incoming plane wave may then be focused to an array of points
in a plane near the surface of the wafer, to produce a tight
optical trap above each reflector. Once integrated into an atom
chip \cite{Fortagh07}, these traps may be loaded either directly
from a surface magneto-optical trap, or from a magnetic guide with
optical molasses added to enable the blockade.

In calculating the characteristics of these microtraps, we found
it necessary to consider both the diffraction and the spherical
aberrations. This led us to derive an analytical expression for
the distribution of the light near the optic axis of a spherical
mirror, which is applicable for wide-angle reflectors and does not
seem to have appeared previously in the literature. Although the
full problem can be solved numerically with a high-performance
computer and sufficient patience, our formula has the advantage
that it can be evaluated more or less instantaneously. In the rest
of this paper, we derive our analytical results, compare them with
full numerical simulations, and show that the silicon
micro-mirrors are indeed suitable for making sub-wavelength atom
traps.

\section{Theory}

\begin{figure}
\centering\includegraphics[width=0.6\columnwidth]{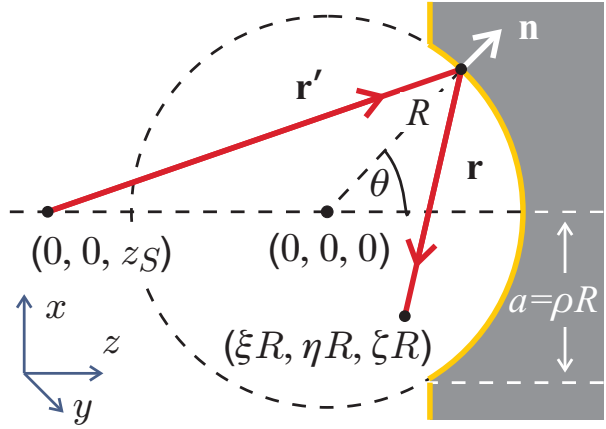}
\caption{\label{fig:geometry}(Color online) Light incident along
the path $\bf{r^\prime}$ is reflected at a point on the spherical
mirror surface, then propagates along $\bf{r}$ to the point of
observation (cartesian coordinates $(\xi R,\eta R,\zeta R)$).}
\end{figure}

Figure \ref{fig:geometry} shows a schematic micro-mirror with
radius of curvature $R$ centered on the point $(0,0,0)$,
illuminated by a point source at position $(0,0,z_{\textrm{s}})$.
We take $z_{\textrm{s}}\rightarrow -\infty$, so that the incident
light is collimated and parallel to the optic axis. If the light
were incident on a simple circular hole of radius $a$, the
amplitude of the output in the far field propagating at angle
$\alpha$ to the optic axis would be given by $J_1(k a\sin \alpha
)/\sin\alpha$, where $k$ is the wavenumber $2\pi/\lambda$ and
$J_1$ is the Bessel function \cite{BornWolf}. As a starting point,
let us consider ideal imaging that satisfies the Abbe
sine criterion and focuses incident collimated light onto the
plane $z=R/2$. In the image plane, the rays incident at angle
$\alpha$ have an off-axis displacement of $(R/2)\tan\alpha$. The
resulting Fraunhofer diffraction pattern in the focal plane has a
radius at half-maximum intensity, $r_{1/2}^{\,\rm diff}$, given by

\begin{equation}\label{eq:diffraction}
\frac{r_{1/2}^{\,\rm diff}}{\lambda} = \frac{1.62\,R/2}{2\pi\,a}=
\frac{0.13}{\rho} \;,\quad\left(\rho\le\frac{1}{2}\right)\,.
\end{equation}
Here, we have neglected the difference between $\sin\alpha$ and
$\tan\alpha$, since we are only interested in apertures having
$a\gg\lambda$, which produce small angles of diffraction. The
aperture size is conveniently expressed by the dimensionless ratio
$\rho\equiv a/R$. For a spherical mirror, $\rho\le 1$, but for
imaging that satisfies the Abbe criterion, the numerical aperture
reaches its maximum value of 1 when the input radius is equal to
the focal length (see \cite{BornWolf}, \S4.5.1). Thus the maximum
aperture for our ideal optic is $\rho=1/2$, giving a minimum spot
size according to Eq.~(\ref{eq:diffraction}) of $0.26\,\lambda$.
Throughout this paper we use the phrase ``spot size'' to mean
radius at half-maximum intensity.

This result captures the essence of the diffraction but does not
treat the focusing correctly since a large-aperture mirror
exhibits numerous orders of aberration. We therefore adopt
Kirchhoff's diffraction theory in order to estimate the spot size
more accurately. The incident light is assumed to illuminate the
mirror surface uniformly. Secondary waves then radiate to the
observation point ${\bf x}_{\rm O}=(\xi R,\eta R, \zeta R)$,
illustrated in Fig.\,\ref{fig:geometry}. For simplicity we ignore
the incoming beam and reflections from the flat part of the mirror
because we are interested in the focal region, where the reflected
intensity is relatively high. Kirchhoff's integral for the
diffracted wave $\psi_\perp^{(0)}$ is then
\begin{eqnarray}\label{eq:kirchhoff}
\psi_\perp^{(0)}({\bf x}_{\rm O}) &=& \frac{1}{4\pi}\int\!\!\!\int
\left[ \frac{e^{ikr^\prime}}{r^\prime}\nabla_{\bf n}G(r) -
G(r)\nabla_{\bf n}\frac{e^{ikr^\prime}}{r^\prime} \right]
d\mathcal{A}\;,
\end{eqnarray}
where $d\mathcal{A}=R^2d\phi\,d\!\cos\theta$ is an element of area
on the curved surface of the mirror. The scalar $r^\prime$ is the
modulus of the vector ${\bf r}^\prime$ shown in
Fig.\,\ref{fig:geometry}, and similarly for $r$. The free-space
Green function for the light field is \mbox{$G(r)=\exp(ikr)/r$},
and $\nabla_{\bf n}$ denotes the gradient taken along $\bf{n}$,
the normal to the mirror surface. Since the observation points of
interest will be many wavelengths away from the mirror surface, we
can assume that $kr\gg1$ and therefore
\begin{eqnarray}\label{eq:nabla}
\nabla_{\bf n}\,\frac{e^{ikr}}{r} &\approx& ik \cos({\bf n},{\bf
r})\,\frac{e^{ikr}}{r}\;,
\end{eqnarray} where $\cos({\bf n},{\bf r})$ is the cosine of the
angle between ${\bf r}$ and ${\bf n}$. A similar expression holds
for $r\rightarrow r^\prime$.

Taking $z_{\rm S}\to-\infty$, the source term
\mbox{$\exp(ikr^\prime)/r^\prime\to\psi\exp(ikR\cos\theta)$}, with
the complex amplitude $\psi$ being constant. Then
\begin{eqnarray}\label{eq:kirchhoff2}
\psi_\perp^{(0)} &=& \frac{ik\,\psi}{4\pi}\int\!\!\!\int
\frac{e^{ik(r + R\cos\theta)}}{r}\left[\cos({\bf n},{\bf r}) -
\cos({\bf n},{\bf r^\prime})\right] d\mathcal{A}\;.
\end{eqnarray}
With this collimated input, $\cos(\bf{n},\bf{r}^\prime)$ is just
$\cos\theta$, while $\cos(\bf{n},\bf{r})$ is given by
\begin{eqnarray}
r\cos({\bf n},{\bf r})= -R\,(1-\xi\cos\phi\sin\theta
-\zeta\cos\theta)\;,
\end{eqnarray}
where we have taken $\eta=0$ without loss of generality.  Since
the light is to be tightly focused near the optical axis, we need
only consider observation points where \mbox{$\xi\ll 1$}.
Expanding $r$ to first order around $\xi=0$, we obtain
\begin{eqnarray}\label{eq:delta}\nonumber
\frac{r}{R} &\approx& \sqrt{1+\zeta^2-2\,\zeta\cos\theta} -
\frac{\xi\,\sin\theta\cos\phi}{\sqrt{1+\zeta^2-2\,\zeta\cos\theta}}
\\ &\equiv& \delta_0 + \xi\,\delta_1\cos\phi\;.
\end{eqnarray}
Returning now to Eq.~(\ref{eq:kirchhoff2}), we eliminate the two
cosines and replace $r$ by  $R\,\delta_0$ everywhere except in the
exponent. There, we keep the $R\,\xi\,\delta_1$ term as well and
integrate over $\phi$, giving
\begin{eqnarray}\label{eq:fullform}\nonumber
\psi_\perp^{(0)} &=& \frac{k\,\psi}{4\pi i}\int\!\!\!\int
\frac{e^{i\kappa(\mu+\delta_0+\xi\delta_1\cos\phi)}}{R\,\delta_0}
\left(\mu+ \frac{1-\zeta\mu}{\delta_0} \right)
d\mathcal{A} \\
&=& \frac{\kappa\,\psi}{2i}\int_{\mu_0}^1
\frac{e^{i\kappa(\mu+\delta_0)}}{\delta_0} \left(\mu +
\frac{1-\zeta\mu}{\delta_0} \right)
J_0\big(\frac{\kappa\,\xi}{\delta_0}\sqrt{1-\mu^2}\:\big)\,d\mu\;,
\end{eqnarray}
where $\kappa=k R$, $\mu=\cos\theta$ and $\mu_0=(1-\rho^2)^{1/2}$.
This is our main result.

In the limit of small mirror aperture, \emph{i.e.} small $\rho$
and $\theta$, $\delta_0 \rightarrow (1-\zeta)$, $\mu \rightarrow
1$, $(1-\mu^2)^{1/2}\rightarrow \theta$ and $d\mu\rightarrow
-\theta d\theta$. Then
\begin{equation}\label{eq:smallaperture}
\left|\psi_\perp^{(0)}\right| \rightarrow \left|\psi\right|
\left(\frac{\rho\, q}{2}\right)\left|\frac{2J_{1}(q\,\xi)}{q\,
\xi}\right|, \quad q=\frac{\kappa\,\rho}{1-\zeta}\;.
\end{equation}
Thus the radius at half-maximum intensity is
\begin{equation}\label{eq:rhalf}
r_{1/2}=0.26\,(1-\zeta)\,\lambda/\rho\quad (\mbox{when } \rho\ll
1)\;.
\end{equation}
Evaluated in the focal plane $\zeta=1/2$, this duplicates
Eq.~(\ref{eq:diffraction}), as one would expect in this limit
where the mirror becomes an ideal thin element. The position of
maximum intensity lies on the $z$ axis, but not at $\zeta=1/2$. In
this small aperture limit, the intensity increases as the mirror
is approached because the spreading of the light due to
diffraction dominates over the focusing due to the curvature of
the mirror surface: this is the limit of small Fresnel number.

\section{Results}

\begin{figure}
\centering\includegraphics[width=0.9\columnwidth]{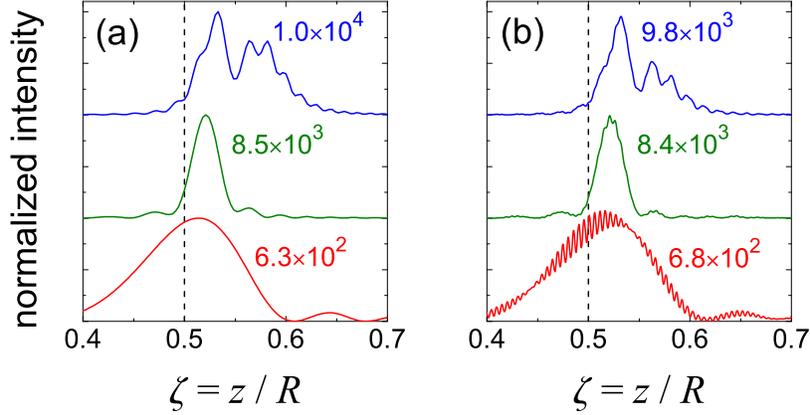}
\caption{\label{fig:cuts}(Color online) Intensity distribution on
axis with $R=100\,\lambda$ for values of mirror aperture
$\rho=0.2$ (bottom), $0.4$ (middle), $0.6$ (top).  These are
offset vertically for clarity. Labels give the peak intensity for
unit incident intensity. (a) Prediction of
Eq.~(\ref{eq:fullform}). Curves are re-scaled to be equal in
height. (b) Result of full numerical integration of Maxwell's
equations. Each curve has the same scale as the corresponding
curve in (a). Dashed lines mark the geometrical focus at
$\zeta=1/2$. }
\end{figure}

\begin{figure}
\centering\includegraphics[width=0.6\columnwidth]{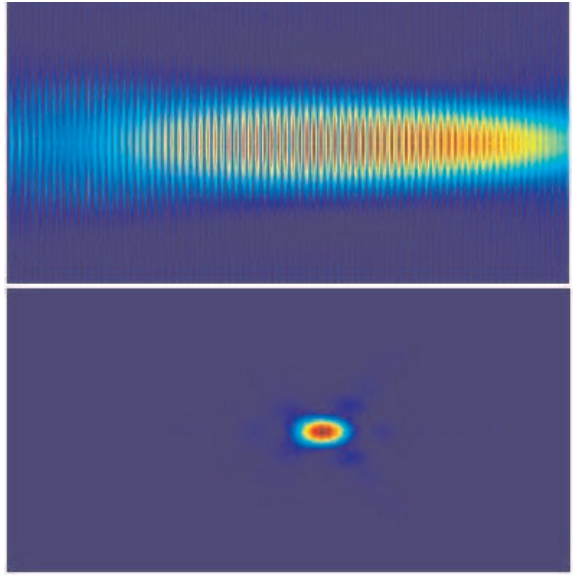}
\caption{\label{fig:MEEP}(Color online) Simulated intensity
distributions of light outside concave spherical mirrors with
radius of curvature $R=100\,\lambda$, etched in a plane substrate.
The field of view is $10\lambda$ (vertical) $\times 40\lambda$
(horizontal), centered on $z=R/2$, with the mirror to the right.
The calculations are done by numerical integration of Maxwell's
equations. Upper image: $\rho=0.1$. Lower image: $\rho=0.4$.}
\end{figure}

Figure~\ref{fig:cuts}(a) shows the intensity on axis versus
position, as given by Eq.~(\ref{eq:fullform}) for three larger
values of the aperture. These all have Fresnel numbers greater
than unity and exhibit peaks close to $\zeta=0.5$. At first, as
the aperture is increased, the light becomes more concentrated
axially, giving rise to a narrower peak at $\rho=0.4$ (middle)
than at $\rho=0.2$ (bottom). With even larger aperture, however,
secondary structure appears on the large-$\zeta$ side of the main
peak. This is due to spherical aberration, \emph{i.e.} to rays
that are incident increasingly far from the axis and therefore
cross the axis at larger values of $\zeta$ after reflection. In
order to test these detailed predictions of
Eq.~(\ref{eq:fullform}), we have integrated Maxwell's equations
numerically using freely available software \cite{MEEP}, based on
the finite-difference time-domain method \cite{FDTD}, with
sub-pixel smoothing for accuracy at sharp interfaces
\cite{Farjadpour06}. Our resolution varies from 20 to 32 pixels
per wavelength and we treat the mirror surface as a perfect
conductor. We exploit the cylindrical symmetry of the problem by
taking the incident plane-wave beam to be circularly polarized.
Figure~\ref{fig:MEEP} illustrates the solutions obtained in this
way for $\rho=0.1$ (upper frame) and $\rho=0.4$ (lower frame). The
numerical results for the intensity on axis are plotted in
Fig.~\ref{fig:cuts}(b) for comparison with the plots shown in
Fig.~\ref{fig:cuts}(a).

For the smallest aperture, there is good agreement on the shape
and intensity of the curve, but the numerical integration exhibits
additional rapidly oscillating fringes, similar to those seen in
the upper frame of Fig.~\ref{fig:MEEP}. These are due mainly to
interference between the incident plane-wave and the field
reflected from the mirror. The same effect can be captured
qualitatively by adding the incident field to
Eq.~(\ref{eq:fullform}). Such an intensity distribution could be
useful for making a tightly confining optical lattice but would
not be chosen when a single, well-defined atom trap is required.
The full numerical solution with medium aperture, shown in the
lower frame of Fig.~\ref{fig:MEEP} and in the central curve of
Fig.~\ref{fig:cuts}(b), agrees very well with our analytical
result. The interference fringes are much less evident here, as
the amplitude of the reflected wave is much higher. There is also
good agreement in the case of the largest aperture, except that
the subsidiary structure due to spherical aberration is slightly
smaller in the exact solution. This is because the electric fields
of rays coming from the outer parts of the mirror are
significantly inclined and should be added as vectors, whereas
Eq.~(\ref{eq:fullform}), based on the scalar wave equation, adds
them as scalars.  We conclude that an aperture $\rho =0.4$ is a
good choice for achieving a single optical dipole trap with tight
axial confinement, and that this is well described by
Eq.~(\ref{eq:fullform}). Note that the use of a larger aperture
does not add substantially to the peak intensity.

\begin{figure}
\centering\includegraphics[width=0.7\columnwidth]{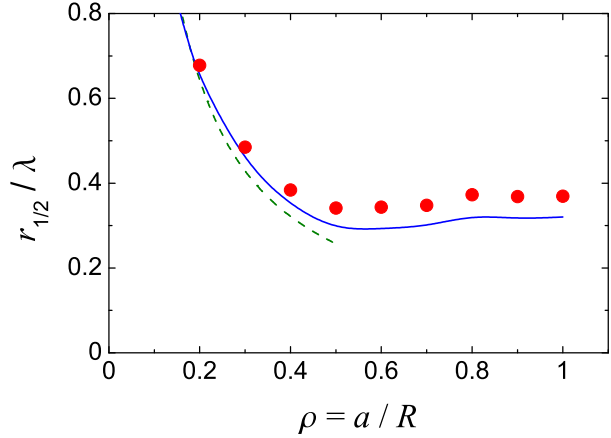}
\caption{\label{fig:varyrho}(Color online) Normalized spot size
$r_{1/2}/\lambda$ as a function of aperture $\rho$ with
$R=100\,\lambda$. Solid line: radius given by
Eq.~(\ref{eq:fullform}) evaluated at the peak of the axial
intensity distribution. Dashed line: radius given by
Eq.~(\ref{eq:diffraction}) for an ideal optic with
focal length $R/2$. Dots: full numerical integration of Maxwell's
equations.}
\end{figure}

We turn now to the radial width of the intensity distribution. The
dashed curve in Fig.~\ref{fig:varyrho} shows the ideal value
$r_{1/2}^{\,\rm diff}$, given by Eq.~(\ref{eq:diffraction}),
plotted against the aperture $\rho$ up to the maximum value
possible, $\rho=1/2$. This suggests that apertures $\rho > 0.2$
should produce very small spot sizes. The solid line, derived from
Eq.~(\ref{eq:fullform}), shows that although the spot size is
somewhat increased by the mirror depth and the aberrations
neglected in Eq.~(\ref{eq:diffraction}), it is still well below
$\lambda/2$. We calculate this width not in the focal plane
$\zeta=1/2$ but at the value of $\zeta$ where the on-axis
intensity is maximum, since that is where the optical dipole trap
is actually formed. The dots show the widths obtained from the
full numerical integration of Maxwell's equations. Near
$\rho=0.2$, these agree closely with the widths derived from
Eq.~(\ref{eq:fullform}), but as the aperture opens, we see that
the full solution gives a slightly larger spot size. This
broadening is another manifestation of the vector nature of the
light field. We conclude that parallel light, incident on a
spherical micro-mirror with aperture $\rho=0.4$, can produce an
optical dipole trap that is well described by
Eq.~(\ref{eq:fullform}), with spot size $<\lambda/2$.

Concerning the vector nature of the field, this is transverse on
axis and follows the polarisation of the input light. Off axis,
however, the field acquires a $z$-component, given in the case of
small $\rho$ by \cite{Lax75,Agrawal83}
\begin{equation}\label{eq:psi1}
\psi_z^{(1)}(\xi,\zeta) =
\frac{i}{\kappa}\,\frac{\partial}{\partial\xi}\,\left[\psi_\perp^{(0)}
(\xi,\zeta)\right]\quad.
\end{equation}
This produces the increase in $r_{1/2}$ that we have noted above.
We can estimate the size of the $z$-component from the derivative
of Eq.~(\ref{eq:smallaperture}), even when the mirror is not
strictly in the small $\rho$ limit. Near the focal plane and to
first order in $\xi$, this gives
$\left|\psi_z^{(1)}/\psi_{\perp}^{(0)}\right|\simeq 2\pi\rho^2
x/\lambda$. For $\rho = 0.4$ this ratio grows from zero on axis to
$0.4$ at $x=r_{1/2}=0.4\,\lambda$. Thus, the trapped atom will not
see a constant polarization as it explores the volume of the trap,
but assuming that its kinetic energy is significantly less than
half the trap depth, the polarization will be approximately that
of the input light.

\begin{figure}[h]
\centering\includegraphics[width=0.7\columnwidth]{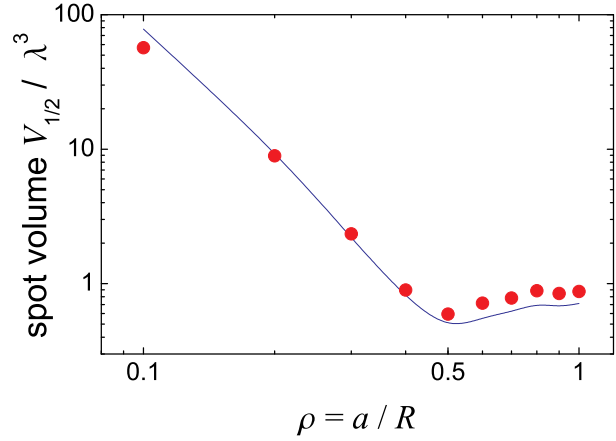}
\caption{\label{fig:volvsrho}(Color online) Normalized spot
volume, as defined in the text, for varying aperture sizes and
$R=100\,\lambda$. Dots: numerical integration of Maxwell's
equations.  Line: results from Eq.~(\ref{eq:fullform}).}
\end{figure}

For collisional blockade experiments, the focal spot must have a
small \emph{volume}, in addition to small $r_{1/2}$, in order to
ensure no more than a single atom per trap. We define the spot
volume $V_{1/2}$ as the volume within the contour of half-maximum
intensity, which is plotted in Figure~\ref{fig:volvsrho} as a
function of $\rho$. For small apertures, we find $V_{1/2}\propto
r_{1/2}^4/\lambda$, as expected from dimensional arguments. For
larger apertures, $V_{1/2}$ drops to a minimum near $\rho=0.5$,
then rises to a level just below $\lambda^{3}$.  As a point of
reference, the spot volume for an ideal Gaussian TEM$_{00}$ beam
is given by $V_{1/2}=41.8\,r_{1/2}^4/\lambda$, which is
$2.6\,\lambda^3$ for $r_{1/2}=\lambda/2$. Once again,
Eq.~(\ref{eq:fullform}) provides a very good approximation to the
exact solution. Finally we note that although all the simulations
shown in this article have taken $R=100\,\lambda$, this is not a
critical requirement. When $R$ is increased from $60\,\lambda$ to
$140\,\lambda$, the spot volume obtained by numerical integration
for $\rho=0.4$ grows by less than $14\%$, though of course the
constraint on surface quality of the mirror becomes more
demanding.

\section{Conclusion}

In conclusion we have derived a useful formula for the field
distribution in front of a spherical mirror illuminated by a plane
wave.  Using this result we have shown that a modern
micro-fabricated mirror of order $10$-$100\,\lambda$ in size can
produce either a tight optical lattice, or a single spot with
little additional structure, having a radius at half-maximum
intensity well below $\lambda/2$. The fabrication method makes it
straightforward to scale this up to a large number of spots in any
desired array. This method is therefore suitable for building atom
traps for applications in quantum information processing, or for
any other application where tight focusing is required and the
low-intensity wings of the spot are unimportant.

\section*{Acknowledgements}
This work was supported by the Royal Society, by EPSRC projects
``Cold Atoms in Microtraps"  and QIPIRC, and by the European
Commission projects SCALA and HIP.  We thank the developers of
MEEP at MIT for providing the free simulation package used in this
work.


\begin{thebibliography}{99}

\bibitem{Schlosser01}
N. Schlosser, G. Reymond, I. Protsenko, and  P. Grangier,
``Sub-poissonian loading of single atoms in a microscopic dipole
trap,'' \nat {\bf 411}, 1024-1027 (2001).

\bibitem{Schlosser02}
N. Schlosser, G. Reymond, and P. Grangier, ``Collisional Blockade
in Microscopic Optical Dipole Traps,'' \prl {\bf 89}, 023005
(2002).

\bibitem{Sortais07}
Y. R. P. Sortais, H. Marion, C. Tuchendler, A. M. Lance, M.
Lamare, P. Fournet, C. Armellin, R. Mercier, G. Messin, A.
Browaeys, and P. Grangier, ``Diffraction-limited optics for
single-atom manipulation,'' \pra {\bf 75}, 013406 (2007).

\bibitem{Darquie05}
B. Darqui\'{e}, M. P. A. Jones, J. Dingjan, J. Beugnon, S.
Bergamini, Y. Sortais, G. Messin, A. Browaeys, P. Grangier,
``Controlled Single-Photon Emission from a Single Trapped
Two-Level Atom,'' Science {\bf 309}, 454-456 (2005).

\bibitem{Weber06}
M. Weber, J. Volz, K. Saucke, C. Kurtsiefer, and H. Weinfurter,
``Analysis of a single-atom dipole trap,'' \pra {\bf 73}, 043406
(2006).

\bibitem{Tey08}
M. K. Tey, Z. Chen, S. A. Aljunid, B. Chng, F. Huber, G.
Maslennikov, C. Kurtsiefer, ``Strong interaction between light and
a single trapped atom without a cavity,'' arXiv:0802.3005v3
(2008), \url{http://uk.arxiv.org/abs/0802.3005v3}.

\bibitem{Lounis05}
B. Lounis and M. Orrit, ``Single-photon sources,'' Rep. Prog.
Phys. {\bf 68}, 1129-1179 (2005).

\bibitem{Waks02}
E. Waks, C. Santori, and Y. Yamamoto, ``Security aspects of
quantum key distribution with sub-Poisson light,'' \pra {\bf 66},
042315 (2002).

\bibitem{Knill01}
E. Knill, R. Laflamme, and G. J. Milburn, ``A scheme for efficient
quantum computation with linear optics,'' \nat {\bf 409}, 46-52
(2001).

\bibitem{Oxborrow05}
M. Oxborrow and A. G. Sinclair, ``Single-photon sources,''
Contemp. Phys. {\bf 46}, 173-206 (2005).

\bibitem{Sondermann07}
M. Sondermann, R. Maiwald, H. Konermann, N. Lindlein, U. Peschel,
and G. Leuchs, ``Design of a mode converter for efficient
light-atom coupling in free space,'' Appl. Phys. B {\bf 89},
489-482 (2007).

\bibitem{vanEnk01}
S. J. van Enk and H. J. Kimble, ``Strongly focused light beams
interacting with single atoms in free space,'' \pra {\bf 63},
023809 (2001).

\bibitem{Tey08b}
M. K. Tey, S. A. Aljunid, F. Huber, B. Chng, Z. Chen, G.
Maslennikov, and C. Kurtsiefer, ``Interfacing light and single
atoms with a lens,'' arXiv:0804.4861v2 (2008),
\url{http://uk.arxiv.org/abs/0804.4861v2}.

\bibitem{Eriksson05}
S. Eriksson, M. Trupke, H. F. Powell, D. Sahagun, C. D. J.
Sinclair, E. A. Curtis, B. E. Sauer, E. A. Hinds, Z. Moktadir, C.
O. Gollasch and M. Kraft, ``Integrated optical components on atom
chips,'' Eur. Phys. J. D {\bf 35}, 135-139 (2005).

\bibitem{Trupke05}
M. Trupke, E. A. Hinds, S. Eriksson, E. A. Curtis, Z. Moktadir, E.
Kukharenka, and M. Kraft, ``Microfabricated high-finesse optical
cavity with open access and small volume,'' \apl {\bf 87}, 211106
(2005).

\bibitem{Fortagh07}
For a recent review of atom chip experiments, see J. Fort\'{a}gh
and C. Zimmermann, ``Magnetic microtraps for ultracold atoms,''
\rmp {\bf 79}, 235 (2007).

\bibitem{BornWolf} M. Born and E. Wolf, {\it
Principles of Optics}, $7^{th}$ edition, (Cambridge University Press, Cambridge,
1999).

\bibitem{MEEP} The software used in this work is available from
\url{http://ab-initio.mit.edu/wiki/index.php/Meep}.

\bibitem{FDTD}
A. Taflove and S. C. Hagness, {\it Computational Electrodynamics:
The Finite-Difference Time-Domain Method}, (Artech, Norwood, MA,
2000).

\bibitem{Farjadpour06}
A. Farjadpour, D. Roundy, A. Rodriguez, M. Ibanescu, P. Bermel, J.
D. Joannopoulos, S. G. Johnson, and G. Burr, ``Improving accuracy
by subpixel smoothing in the finite-difference time domain,'' \ol
{\bf 31}, 2972-2974 (2006).

\bibitem{Lax75}
M. Lax, W. H. Louisell, and W. B. McKnight, ``From Maxwell to
paraxial wave optics,'' \pra {\bf 11}, 1365-1370 (1975).

\bibitem{Agrawal83}
G. P. Agrawal and M. Lax, ``Free-space wave propagation beyond the
paraxial approximation,'' \pra {\bf 27}, 1693-1695 (1983).

\end{thebibliography}
\end{document}